\documentclass[twocolumn,prc,superscriptaddress]{revtex4}
\usepackage{epsfig}
\usepackage{bm}

\begin{document}
\newcommand{\nl}{\nonumber \\}
\newcommand{\bea}{\begin{eqnarray}}
\newcommand{\eea}{\end{eqnarray}}
\newcommand{\bi}{\bibitem}
\newcommand{\be}{\begin{equation}}
\newcommand{\ee}{\end{equation}}
\newcommand{\bt}{\begin{table}}
\newcommand{\et}{\end{table}}
\newcommand{\btab}{\begin{tabular}}
\newcommand{\etab}{\end{tabular}}
\newcommand*{\rf}[1]{(\ref{#1})}
\newcommand*{\opava}{Institute of Physics, Silesian University in Opava,
Bezru\v{c}ovo n\'{a}m. 13, 746 01 Opava, Czech Republic}
\newcommand*{\praha}{Institute of Experimental and Applied Physics,
Czech Technical University, Horsk\'{a} 3/a, 120 00 Prague, Czech Republic}
\def\rms{M_\rho^2(s)}
\def\mrs{m_{\rho}^{2}}
\def\mgs{m_\rho\Gamma_\rho(s)}
\def\mps{m_\pi^{2}}
\def\grs{\Gamma_\rho(s)}
\def\ra{\rightarrow}
\def\dek#1{\times\!10^{#1}}
\def\die{$e^+e^-$}
\def\rpp{\rho^0\ra\pi^+\pi^-}
\def\op{\omega\pi^0}
\def\kk{ K\bar K}
\def\eepp{e^+e^-\rightarrow\pi^+\pi^-}
\def\eg{e.g.}
%
% Journal and other miscellaneous abbreviations for references
\def \zpc#1#2#3{Z.~Phys.~C \textbf{#1}, #2 (#3)}
\def \plb#1#2#3{Phys.~Lett. B \textbf{#1}, #2 (#3)}
\def \prl#1#2#3{Phys.~Rev.~Lett. \textbf{#1}, #2 (#3)}
\def \prpt#1#2#3{Phys.~Rep. \textbf{#1}, #2 (#3)}
\def \prc#1#2#3{Phys.~Rev.~C \textbf{#1}, #2 (#3)}
\def \prd#1#2#3{Phys.~Rev.~D \textbf{#1}, #2 (#3)}
\def \prev#1#2#3{Phys.~Rev. \textbf{#1}, #2 (#3)}
\def \npa#1#2#3{Nucl.~Phys.~A \textbf{#1}, #2 (#3)}
\def \nca#1#2#3{Nuovo Cimento~A \textbf{#1}, #2 (#3)}
\def \npb#1#2#3{Nucl.~Phys.~B \textbf{#1}, #2 (#3)}
\def \rmp#1#2#3{Rev.~Mod.~Phys. \textbf{#1}, #2 (#3)}
\def \jpg#1#2#3{J.~Phys. G \textbf{#1}, #2 (#3)}
\def \ea{\textit{et~al.}}
\def \ibid#1#2#3{\textit{ibid.} \textbf{#1}, #2 (#3)}
\def \sjnp#1#2#3#4{Yad. Fiz. \textbf{#1}, #2 (#3) [Sov. J. Nucl. Phys.
\textbf{#1}, #4 (#3)]}
\def \jetp#1#2#3#4#5{Zh. Eksp. Teor. Fiz. \textbf{#1}, #2 (#3) 
[J. Exp. Theor. Phys. \textbf{#4}, #5 (#3)]}
\def \epja#1#2#3{Eur. Phys. J. A \textbf{#1}, #2 (#3)}
\def \epjc#1#2#3{Eur. Phys. J. C \textbf{#1}, #2 (#3)}
\def \jetpl#1#2#3#4#5{Pisma v Zh. Eksp. Teor. Fiz. \textbf{#1}, #2 (#3)
[JETP Lett. \textbf{#4}, #5 (#3)]}
\def \aps#1#2#3{Acta Phys. Slovaca \textbf{#1}, #2 (#3)}

\title{An Alternative Parametrization of the Pion Form Factor\\
and the Mass and Width of \bm{$\rho$}(770)}

\thanks{This paper is dedicated to the late Julia Thompson, 
who drew the attention of one of us (P.~L.) to the experimental program 
of the Budker Institute of Nuclear Physics at Novosibirsk.}

\author{Peter Lichard}
\affiliation{\opava}
\affiliation{\praha}
\author{Martin Voj\'{i}k}
\affiliation{\opava}

\date{\today}
\begin{abstract}
In order to reveal possible mass shifts of the vector mesons in a
dense strongly interacting medium presumably created in high energy
heavy ion collisions it is necessary to know their free masses reliably. 
The $\rho$(770) mass quoted in the last two editions of the
Review of Particle Physics is significantly larger than the values 
quoted in previous editions. The new value is mostly influenced by the
results of recent experiments CMD-2 and SND at the VEPP-2M \die collider at
Novosibirsk. We show that the values of the mass and width of the $\rho$(770) 
meson measured in the $\eepp$ annihilation depend crucially on the 
parametrization used for the pion form factor. We propose a parametrization 
of the $\rho$(770) contribution to the pion form factor based on the 
running mass calculated from a single-subtracted dispersion relation
and compare it with the parametrization based on the formula of Gounaris
and Sakurai used recently by the CMD-2 collaboration. We show that our
parametrization gives equally good or better fits when applied to the
data of the CMD-2, SND, and KLOE collaborations, but yields much smaller 
values of the $\rho$(770) mass, consistent with the photoproduction 
and hadronic reactions results. Our fit to the KLOE data becomes
exceptionally excellent (confidence level of 99.88\%) if an energy shift 
of about 2~MeV in the $\rho$-$\omega$ region is allowed. 
\end{abstract}

\maketitle

\section{\label{sec:intro}Introduction}
The vector meson $\rho(770)$ plays an important role in many phenomena of
particle and nuclear physics. Possible modification of its mass and width 
in dense strongly interacting systems presumably created in high energy 
heavy ion collisions is now widely discussed
\cite{barz1991,eletzky98,rapp2000,brown2002,gale02,kolb2003,pratt2003,%
shuryak2003,rapp2003,brown2004,ruppert06}.
To study this effect quantitatively, reliable information about the
``vacuum" values of these parameters is desirable. 

Inspecting recent editions of the Review of Particle Physics 
\cite{pdg1992,pdg1994,pdg1996,pdg1998,pdg2000,pdg2002,pdg2004,pdg2006} 
we find: 
(i) whereas
the values of $\rho(770)$ mass quoted until 2002 are compatible
within experimental errors, an abrupt rise appears in 2004 (see Table
\ref{tab1}); (ii) the $\rho^0$ mass average calculated from the 
\die~annihilation experiments is much higher
than the averages calculated from hadronic reactions and photoproduction.
The figures taken from the 2006 Particle Date Group (PDG) tables are shown 
in Table \ref{tab2}.
\begin{table}
\caption{\label{tab1}Mass and width of $\rho(770)$ according to PDG}
\begin{ruledtabular}
%\begin{tabular}{lll}
\begin{tabular}{ccc}
 year & $m_{\rho}$ & $\Gamma_{\rho}$\\
      & (MeV)             & (MeV)\\
 \hline
 1992 \cite{pdg1992}& $768.1\pm0.5$ & $151.5\pm1.2$\\
 1994 \cite{pdg1994}& $769.9\pm0.8$ & $151.2\pm1.2$\\
 1996 \cite{pdg1996}& $768.5\pm0.6$ & $150.7\pm1.2$\\
 1998 \cite{pdg1998}& $770.0\pm0.8$ & $150.7\pm1.1$\\
 2000 \cite{pdg2000}& $769.3\pm0.8$ & $150.2\pm0.8$\\
 2002 \cite{pdg2002}& $771.1\pm0.9$ & $149.2\pm0.7$\\
 2004 \cite{pdg2004}& $775.8\pm0.5$ & $150.3\pm1.6$\\
 2006 \cite{pdg2006}& $775.5\pm0.4$ & $149.4\pm1.0$\\
\end{tabular}
\end{ruledtabular}
\end{table}

\begin{table}
\caption{\label{tab2}Mass of $\rho$ measured in different processes,
selected and averaged by PDG in 2006 \cite{pdg2006}}
\begin{ruledtabular}
%\begin{tabular}{lll}
\begin{tabular}{lcc}
      & $m_{\rho^0}$ & $m_{\rho^\pm}$\\
      & (MeV)        & (MeV) \\
 \hline
 $\eepp$, $\pi^+\pi^-\pi^0$ & $775.5\pm0.4$ &
$775.4\pm0.4$\footnote{Includes also $\tau$ decay results} \\
 Photoproduction & $768.5\pm1.1$ & / \\
 Hadronic reactions & $769.0\pm0.9$ & $766.5\pm1.1$ \\
\end{tabular}
\end{ruledtabular}
\end{table}

The discrepancy between the \die~annihilation measurements of the 
$\rho^0$ mass and the results of photoproduction and hadronic 
experiments is very troublesome. We know of no single theoretical 
prediction or explanation of such a phenomenon. 

In the literature one can find many articles that dealt with
the determination of the $\rho(770)$ parameters from the experimental data
on the pion form factor 
\cite{heyn81,geshkenbein84,geshkenbein89,dubnicka89,bernicha94,connell97,%
gardner98,benayoun98,benayoun99,dubnicka99,melikhov04}. Their results on 
the $\rho^0$ mass are smaller, in some cases substantially, than the value 
preferred now by the PDG. It is not clear what has led the PDG to not 
using them ``for averages, fits, limits, etc.'' \cite{pdg2006}. The 
PDG average builds on five experiments, three of which explored
the \die~annihilation into a pair of charged pions. Two of those three
experiments utilized the Gounaris-Sakurai (GS) parametrization \cite{GS}
when fitting the data. 

As we discuss in detail in the next section, 
the validity of the GS formula is limited because it ignores two important
effects. Namely, the opening of additional decay channels at higher energies
and the structure effects in the hadronic vertices manifesting themselves 
as the momentum dependent strong form factors. These effects directly 
influence the imaginary part in the denominator of the GS formula. This 
itself would not have serious implications for the analysis of the form 
factor data, which are usually restricted to an narrow interval around the
nominal $\rho$ mass. But because of the close connection between the 
denominator's  real and imaginary part implied by the analyticity, the
real part is also influenced, even at the energies around the nominal $\rho$
mass, which are important for data fitting.

Already in the pioneer era of the $\rho$-meson physics, when the $\rho$ 
mass and width were determined mainly by analyzing the dipion mass 
distribution in the reaction $\pi p\ra\pi\pi N$, Pi\v{s}\'{u}t and 
Roos \cite{pisut68} showed that the discrepancies among the results 
of different experimental groups are caused by different parametrizations
these groups used. They also discussed the discrepancy between the
hadronic measurements and the results of the first $e^+e^-$ annihilation
into two pions experiment \cite{auslender67}. From that time the
parametrization dependence, mainly in connection with the pion form
factor parametrization, has been discussed in many papers, see, e.g., 
Refs.~\cite{roos75,bernicha94,gardner98,benayoun99,feuillat01,melikhov04}.

The aim of this paper is to show that the uncertainties connected with
the choice of the form factor parametrization induce much larger errors
of the $\rho$ resonance parameters than the statistical and systematic
errors of the data. For that purpose we fit three different data sets
with two different parametrizations and compare the resulting masses and 
width of the $\rho^0$ meson. One of the parametrizations is that
used recently by the CMD-2 collaboration \cite{cmd2_2002,cmd2_2004}. Its 
principal element 
is the GS formula. The other parametrization \cite{ratio} is based on a
single-subtraction dispersion relation fed by the imaginary part taking
into account several decay channels which open when the $\rho$-resonance
goes ``off-mass-shell'' towards higher masses. All decay widths are 
calculated taking into account the strong form factors. 

The paper is organized as follows. In Sec.~\ref{sec:para} we describe
the two versions of the $\rho^0$ resonance term in the pion
form factor. The parametrization of the subleading contributions coming 
from the $\rho(1450)$, $\omega(782)$, and $\phi(1020)$ is presented
in Sec.~\ref{sec:other}. The results of fitting the CMD-2 \cite{cmd2_2004},
SND \cite{snd2006}, and KLOE \cite{kloe} data sets by the two 
parametrizations are shown and discussed in Sec.~\ref{sec:results}. 
Section \ref{sec:summary} contains summary and concluding remarks.

\section{\label{sec:para}Two parametrizations of the \bm{$\rho(770)$} 
contribution to the pion form factor}
We write the contribution of the intermediate state with the $\rho^0$ meson
to the pion electromagnetic form factor in the following form
\be
\label{piffgen}
F_\rho(s)=\frac{M^2_\rho(0)}{\rms-s-i\mgs},
\ee 
where the running mass squared $\rms$ is a real function satisfying the
conditions
\be
\label{rms_sr}
M^2_\rho(\mrs)=\mrs,
\ee
and
\be
\label{rmsderiv}
\left.\frac{d \rms}{d s}\right|_{s=\mrs}=0,
\ee
$m_\rho$ is the nominal mass of the $\rho$ meson, and
the real function $\grs$ is the total decay width of a $\rho^0$ meson
with mass $\sqrt{s}$. 
It follows from microcausality that the pion form factor $F(s)$ is an 
analytic function
in the whole complex $s$-plane with a cut running from $s=4m^2_\pi$ 
to infinity, satisfying the condition $F(s^*)=F^*(s)$, and growing no faster
than a finite power of $s$ for $|s|\ra\infty$ (in QCD, $|F(s)|$ even
decreases \cite{farrar_etc}). These properties of $F(s)$ imply the analytic 
properties of the denominator in Eq. \rf{piffgen}.

\subsection{\label{sec:gs}Gounaris-Sakurai formula}
Gounaris and Sakurai \cite{GS} assumed a low-energy parametrization
of the phase shift in the $P_{11}$ wave of the $\pi\pi$ elastic scattering
inspired by work of Chew and Mandelstam \cite{chew} and used the relation
between the pion form factor and the $\pi\pi$ elastic scattering amplitude 
in the $P_{11}$ channel \cite{federbush}. After defining the resonance 
parameters in terms of the $P_{11}$ phase shift they obtained 
a parametrization of the pion form factor. Below, we write it in a little 
different form, adopting partly the notation of Ref.~\cite{cmd2_2002} and 
referring to our Eq.~\rf{piffgen}. Firstly, on the upper branch of the cut
we have  
\be
\label{gammags}
\grs=\Gamma_\rho\frac{m_\rho}{\sqrt{s}}
\left(\frac{s-4\mps}{\mrs-4\mps}\right)^{3/2},
\ee
where $\Gamma_\rho=\Gamma_\rho(\mrs)$ is the nominal width of the 
$\rho(770)$ meson, whereas $\grs \equiv 0$ for $s\leq 4\mps$. Secondly,
\be
\label{rmsgs}
\rms=\mrs+f(s),
\ee
where
\bea
f(s)&=&\frac{2\Gamma_\rho\mrs}{(\mrs-4\mps)^{3/2}}\Big\{(s-4\mps)
\big[h(s)-h(\mrs)\big] \nl
&+&(\mrs-4\mps) h^\prime(\mrs)\left(\mrs-s\right)\Big\},
\label{fs}
\eea
\[
h(s)=\left\{
\begin{array}{l}
\frac{1}{\pi}\sqrt{\frac{s-4\mps}{s}}\ln{\frac{\sqrt{s}+\sqrt{s-4\mps}}
{2m_\pi}},\quad s\geq 4 \mps,\\
\\
\sqrt{\frac{4\mps-s}{4s}}\left[1-\frac{2}{\pi}\arctan\sqrt{\frac{4\mps-s}{s}}
\right],\\
\quad 0<s<4\mps,\\
\\
\frac{1}{\pi},\quad s=0,\\
\\
\frac{1}{\pi}\sqrt{\frac{s-4\mps}{s}}\ln{\frac{\sqrt{-s}+\sqrt{4\mps-s}}
{2m_\pi}},\quad s< 0.
\end{array}\right.
\]
To get more insight into the physics behind the GS parametrization let
us utilize that $\rms$ and $\grs$, which are related to the $\rho$-meson
contribution to the pion form 
factor by Eq.~\rf{piffgen}, should be the boundary values of the real 
and imaginary part, respectively, of a function analytic in the cut
$s$-plane. It is easy to check that the quantities \rf{gammags} and 
\rf{rmsgs} are related through the double-subtracted dispersion relation
\be
\label{doubledr}
\rms=A_0+A_1s-\frac{s^2}{\pi}{\mathcal P}\!\!\!\!\int\limits _{4\mps}^\infty
\frac{m_\rho\Gamma_\rho(s^\prime)}{{s^\prime}^2(s^\prime-s)}ds^\prime.
\ee
The appearance of the two subtraction constants $A_0$ and $A_1$ guarantees
that the conditions \rf{rms_sr} and \rf{rmsderiv} can be satisfied.
Of course,
the relation \rf{doubledr} is not specific for the GS parametrization, but 
is more general. It enables us to find the running mass squared $\rms$ 
corresponding to any reasonable (i.e., not growing too fast) choice of 
the total width $\grs$. It is important to realize that the values 
of $\rms$ at any $s$, in particular those for low $s<1$~GeV$^2$, which 
are used in the
form factor fitting, are given by the values of $\grs$ at all $s$.
To get reliable $\rms$ it is therefore essential to have $\grs$ that
is physically sound not only in the $s$-region vital for the fitting, but
also beyond it.

The GS parametrization of the $\rho(770)$ contribution to the pion 
electromagnetic form factor ignores possible structure effects
in the $\rho^0\pi^+\pi^-$ vertex (the strong form factor).
They were taken into account by Vaughn and Wali (VW) \cite{vaughn} in the form
of a phenomenological cutoff function. When that function is set to unity,
the VW parametrization reduces to the GS one. The VW approach has not become
as popular as that of GS, although it is closer to reality, because the outcome 
of dispersion relation \rf{doubledr} cannot be expressed in an analytic
form, similar to \rf{fs}.

A different attempt to bring the GS-type formulas closer to reality was
done by the authors of Ref.~\cite{melikhov04}. They included also the
contributions from the $K^+K^-$ and $K^0\bar{K}^0$ states to the imaginary
part of the denominator of the pion form factor \rf{piffgen} and obtained
its real part in an analytic form using the dispersion relation. No strong
form factor was included. Also this improvement of the GS formula passed
unnoticed by the experimentalists fitting their data. 

We should also note that the original derivation of the GS formula brings 
a one-to-one correspondence between the $P_{11}$ scattering length and 
effective radius on one side, and the $\rho(770)$ resonance position and 
width on the other. Using the GS formula for the contribution of higher 
resonances [$\rho(1450)$, $\rho(1700)$] to the pion form factor, as in 
\cite{barate97,cmd2_2002,cmd2_2004,cmd2_2005}, 
thus contradicts the spirit of the GS derivation as it implies incorrect 
values of the $P_{11}$ scattering length and effective radius. In the
language of dispersion relation \rf{doubledr} it corresponds to the
assumption that the two-pion mode is the only decay mode 
at any $\sqrt{s}$. This is unrealistic because already for the
$\rho(1450)$ with nominal mass there are several other modes allowed
by the energy-momentum, angular momentum, and internal quantum numbers 
conservation. 

\subsection{\label{sec:rm}Our running mass approach}
We build our alternative parametrization on the recent work by one of
us \cite{ratio}, to which the reader is referred for technical details.
The numerical values of the constants used in \cite{ratio} are updated
using the 2002 edition  of the Review of Particle Physics \cite{pdg2002}.

There are four items in which our parametrization differs from the GS 
one:

(i) We consider several contributions to the total decay width of the
$\rho$ meson, which become operational when its mass $\sqrt{s}$ exceeds 
respective thresholds
\bea
\grs&=&\Gamma_{\rho\ra\pi\pi}(s)+\Gamma_{\rho\ra\omega\pi}(s)\nl
&+&
\Gamma_{\rho\ra K\overline{K}}(s)+\Gamma_{\rho\ra\eta\pi\pi}(s).
\label{grsrm}
\eea

(ii) The partial decay width of the two-pion decay, which provides the
essential contribution to \rf{grsrm} is calculated from the standard
interaction Lagrangian containing the first derivative of the pion field.
The result is  
\be
\label{gamrhpp}
\Gamma_{\rho\ra\pi\pi}(s)=\Gamma_\rho\frac{\mrs}{s}
\left(\frac{s-4\mps}{\mrs-4\mps}\right)^{3/2}.
\ee

(iii) The strong form factors are introduced into all purely hadronic 
vertices. They were taken from the flux-tube-breaking model of Kokoski
and Isgur \cite{kokoski}. This leads to the modification of the partial
decay widths composing \rf{grsrm} and calculated originally as if the
participating mesons were elementary quanta. For example, the two-pion
width \rf{gamrhpp} is modified by the following multiplication factor 
\[
\exp\left\{\frac{\mrs-s}{24\beta^2}\right\},
\]
where the value $\beta=0.4$ was taken from the Kokoski and Isgur
\cite{kokoski} analysis.

(iv) As now the the total width \rf{grsrm} does not rise so fast as in the GS
approach, we may write a once-subtracted dispersion relation
\be
\label{singledr}
\rms=B_0-\frac{s}{\pi}{\mathcal P}\!\!\!\!\int\limits _{4\mps}^\infty
\frac{m_\rho\Gamma_\rho(s^\prime)}{{s^\prime}(s^\prime-s)}ds^\prime.
\ee
The subtraction constant $B_0$ is chosen to ensure that the condition
$M^2_\rho(\mrs)=\mrs$ \rf{rms_sr} is satisfied. Now it is not possible
to satisfy the condition on derivative \rf{rmsderiv} automatically.
We consider this an advantage, because Eq.~\rf{rmsderiv}  can thus serve 
as a test that the energy dependent total decay width \rf{grsrm} was 
chosen correctly. For details, see Ref.~\cite{ratio}.
Let us note that our choice of the subtraction point $s=0$ differs from
$s=\mrs$, used in Refs. \cite{ratio,isgur}. The small
disadvantage lying in the necessity to evaluate the subtraction constant
$B_0$ is generously compensated by much faster numerical evaluation of the
dispersion integral. In the following, we are going to vary $m_\rho$ and
$\Gamma_\rho$ in a search for the best fit to the experimental data.
It means that we will have to calculate the running mass \rf{singledr}
in the experimental points again and again. The computation speed is
therefore a decisive factor. 

\begin{figure}
\setlength \epsfxsize{8.6cm}
\epsffile{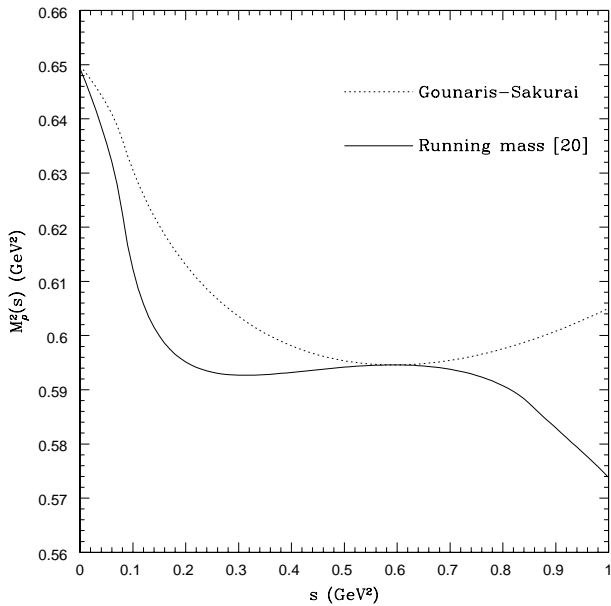}
\caption{\label{fig:compare_rm}Comparison of the running mass squared of 
$\rho$-meson in the Gounaris-Sakurai and our approach}
\end{figure}

\begin{figure}
\setlength \epsfxsize{8.6cm}
\epsffile{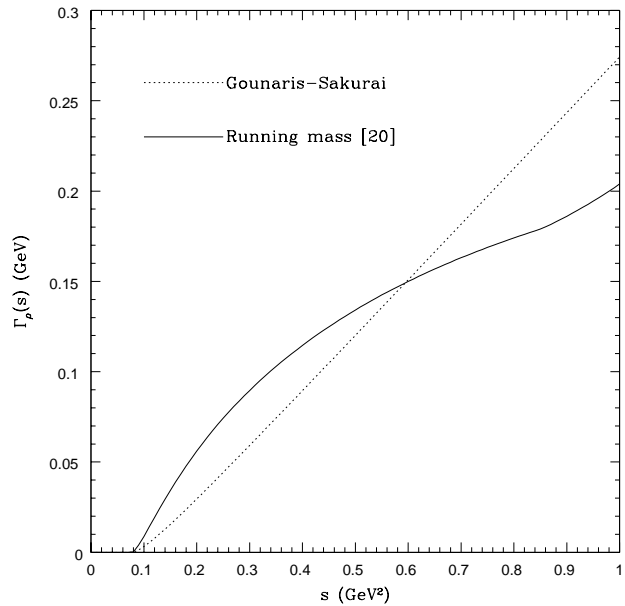}
\caption{\label{fig:compare_gam}Comparison of the energy dependent width of 
$\rho$-meson in the Gounaris-Sakurai and our approach}
\end{figure}

The comparison of the basic quantities of the GS and our approach is 
shown in Fig.~\ref{fig:compare_rm} (running mass) and
Fig.~\ref{fig:compare_gam} (total width). The values
$m_\rho=771.1$~MeV/$c^2$ and $\Gamma_\rho=149.2$~MeV \cite{pdg2002} were 
chosen for this purpose.

\section{\label{sec:other}Other contributions to the pion form factor}

The parametrization of the pion electromagnetic form factor should contain,
in addition to the dominant contribution from the $\rho(770)^0$, also the
contribution from other resonances. In the energy range of the $\eepp$
experiments we are interested in, they include $\rho^\prime\equiv\rho(1450)$, 
$\omega(782)$, and perhaps also $\phi(1020)$. We write the pion form factor 
in the form
\bea
\label{fullpiff}
F(s)&=&\Big\{F_\rho(s)\big[1+\delta\frac{s}{m_\omega^2}F_\omega(s)\big]
+\beta F_{\rho^\prime}(s)+\gamma F_\phi(s)\Big\}\nl
&\times&(1+\beta+\gamma)^{-1},
\eea
where not only $F(s)$ but also the individual contributions to it are 
normalized to unity at $s=0$. Parameters $\beta$, $\gamma$, and $\delta$
are complex and should be determined, together with $m_\rho$ and
$\Gamma_\rho$ by minimizing $\chi^2$ constructed from the data and
parametrizations in the standard way. Equation \rf{fullpiff} is inspired 
by Eq.~(8) from \cite{cmd2_2002} and reduces to it for $\gamma=0$. We 
will use it in conjunction with both GS and our running mass 
parametrizations. In one case (CMD-2 data) allowing non-vanishing $\gamma$ 
did not bring a considerable drop in $\chi^2$. But for the other two data
sets (SND, KLOE) the $\phi$ contribution improved the fits.

The CMD-2 collaboration \cite{cmd2_2002,cmd2_2004} used the GS parametrization 
also for the $\rho(1450)$ contribution and we will do the same when 
applying their formalism. But when using our parametrization, we
utilize the fixed-mass formula
\be
\label{rhoprime}
F_{\rho^\prime}(s)=\frac{m_{\rho^\prime}^2}{m_{\rho^\prime}^2-s-
im_{\rho^\prime}\Gamma_{\rho^\prime}(s)}
\ee
with several options for the total width of $\rho^\prime$: 
\begin{enumerate}
\item $\Gamma_{\rho^\prime}(s)\equiv\Gamma_{\rho(1450)}$;
\item Eq.~\rf{gammags} adapted for $\rho(1450)$;
\item $\Gamma_{\rho^\prime}(s)\equiv 0$;
\item Eq.~\rf{gamrhpp} adapted for $\rho(1450)$.
\end{enumerate}
The options 2 and 4 are equivalent to the assumption that the only
decay mode of $\rho(1450)$ is the dipion one. The option 3 mimics the
situation when $\rho(1450)$ disintegrates only through the modes with
higher thresholds, the decay widths
of which become negligible below $s=1$~GeV$^2$ mainly due to the shrinking 
phase-space. This is in conformity with what was known already a long
time ago, namely that the $\rho^\prime$ resonance is clearly visible
in the $e^+e^-$ annihilation into four pions \cite{conversi74} whereas
its appearance in the two-pion channel is less pronounced, see, e.g., 
Fig. 2 in \cite{cmd2_2005}.

The results of the fits ($\chi^2$, $m_{\rho(770)}$, $\Gamma_{\rho(770)}$)
showed only marginal differences among various options, with option 3
giving slightly better $\chi^2$ than the other options. The results
of our approach presented below were obtained using just that option.

As to the contribution of the $\omega(782)$ to the pion electromagnetic 
form factor, we use
\be
\label{omega}
F_\omega(s)=\frac{m_\omega^2}{m_\omega^2-s-
im_\omega\Gamma_\omega(s)}.
\ee
The following options are considered: 
\begin{enumerate}
\item $\Gamma_\omega(s)\equiv\Gamma_{\omega(782)}$ above the $s=9m_\pi^2$
threshold;
\item  $\Gamma_\omega(s)=\Gamma_{\omega\ra 3\pi}(s)$;
\item same as 2, but including strong form factors;
\item  $\Gamma_\omega(s)=\Gamma_{\omega\ra
3\pi}(s)+\Gamma_{\omega\ra\pi\gamma}(s)+\Gamma_{\omega\ra\pi^+\pi^-}(s)$;
\item same as 4, but including strong form factors.
\end{enumerate}
Following \cite{cmd2_2002}, the option 1 was used in conjunction with the
GS parametrization of the $\rho(770)$ and $\rho(1450)$ contributions. The
other options were explored as a part of our approach, but did not
exhibit notable differences in the quality of the fits.

In the case of the SND data, the contribution of the $\rho$-$\phi$ 
interference is present in a statistically significant way, as judged 
from the $\chi^2$ decrease, with both parametrizations (GS, ours) which 
we explore. For data of the KLOE collaboration, the situation is more 
complicated. The details are reported in Sec.~\ref{sec:kloe}. Our
parametrization of the $\phi(1020)$ contribution to the pion form factor 
is
\be
\label{phi}
F_\phi(s)=\frac{m_\phi^2}{m_\phi^2-s-
im_\phi\Gamma_\phi(s)}.
\ee
We consider the following options: 
\begin{enumerate}
\item $\Gamma_\phi(s)\equiv\Gamma_{\phi(1020)}$ above the $s=4m_K^2$
threshold;
\item  $\Gamma_\phi(s)=\Gamma_{\phi\ra K\overline{K}}(s)$;
\item same as 2, but including strong form factor;
\item  $\Gamma_\phi(s)=\Gamma_{\phi\ra K\overline{K}}(s)
+\Gamma_{\phi\ra 3\pi}(s)+\Gamma_{\phi\ra\eta\gamma}(s)$;
\item same as 4, but including strong form factors.
\end{enumerate}
Given the narrow width of the $\phi$ resonance, the high thresholds
of its important decay modes, and the distance of the experimental energies 
from  the $\phi$ mass, the refinements 2--5 did not change the results 
noticeably in comparison with the simplest option 1.

\section{\label{sec:results}Results}
\subsection{\label{sec:cmd2}Using the CMD-2 experiment data}

In 2002, the CMD-2 Collaboration published \cite{cmd2_2002} their results 
of the precise measurement of the $\eepp$ cross section for 43 c.m. energies
and gave also the values of the pion electromagnetic form factor determined
from them. After an error in the computer code was found, they published 
the updated values of the pion form factor \cite{cmd2_2004}, which we are 
going to explore here. The original paper \cite{cmd2_2002} remains important
as it contains details about the GS based parametrization, which the 
collaboration used in fitting the form factor data.

We first tested our computer code using the uncorrected CMD-2 data because
in Ref.~\cite{cmd2_2002} a more complete information is provided than
in the later paper \cite{cmd2_2004}. Our results agreed with the values
in Table~3 \cite{cmd2_2002}. Then we used the corrected pion form factor
data \cite{cmd2_2004} and got the results shown in Tab.~\ref{tab_cmd2}. 
The mass and width of the $\rho$ meson we obtained
differ only very slightly from the CMD-2 values, differences are much
smaller than the errors quoted in \cite{cmd2_2004}. Our errors shown
in Tab.~\ref{tab_cmd2} and elsewhere are only the statistical ones in the
\textsc{MINUIT} \cite{minuit} sense. 

Our running mass parametrization gives a little better fit to the CMD-2
data than the GS one. The results are shown in the last column of
Tab.~\ref{tab_cmd2}. The $\rho(770)$ mass and width we got are much 
lower that the corresponding quantities obtained by the GS parametrization. 

It is also interesting to compare the mean square radius of the charged
pion, which is related to the derivative of the form factor at $s=0$
\cite{barkov}, with the experimental value $<r_\pi^2>$=(0.439$\pm$0.008)
fm$^2$ \cite{amendolia}. 

\begin{figure}
\setlength \epsfxsize{8.6cm}
\epsffile{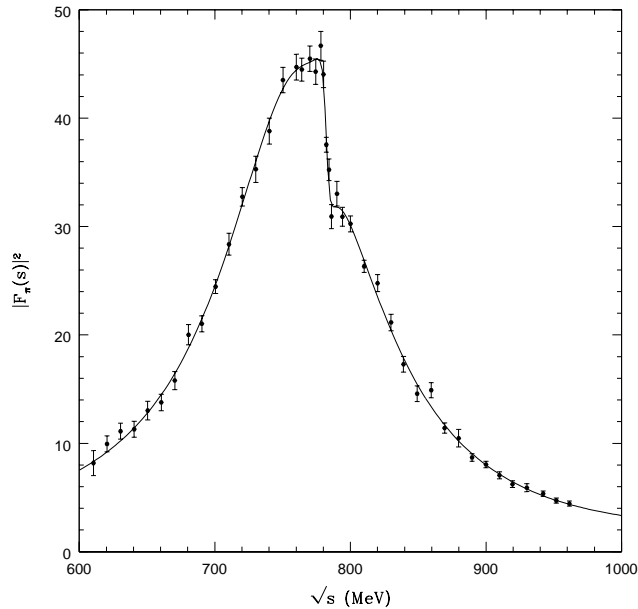}
\caption{\label{fig:cmd2}Comparison of our parametrization of the pion
form factor with the CMD-2 data \cite{cmd2_2004}}
\end{figure}

The graphical comparison of our parametrization with the CMD-2 data
is shown in Fig.~\ref{fig:cmd2}.

\begin{table}
\caption{\label{tab_cmd2}Mass and width of $\rho(770)$ and the mean-square
radius (msr) of the pion from fitting the CMD-2 data. The experimental value 
of the msr is 0.439$\pm$0.008 fm$^2$ \cite{amendolia}.}
\begin{ruledtabular}
\begin{tabular}{lcc}
%\begin{tabular}{ccc}
                 & GS                      & ours\\
 \hline
 $\chi^2/NDF$    & $35.16/37$     &    32.18/37\\
 $m_\rho$~(MeV/$c^2$)      & $775.3\pm1.1$  &$767.08\pm0.83$\\
 $\Gamma_\rho$~(MeV) & $143.2\pm2.5$  &$136.1\pm2.2$  \\
 $<r_\pi^2>$~(fm$^2)$   & $0.416\pm0.013$  &$0.434\pm0.012$\\
\end{tabular}
\end{ruledtabular}
\end{table}

\subsection{\label{sec:snd}Using the SND experiment data}
In 2005, the results were reported \cite{snd2005} of the $\eepp$ cross 
section measurement with the spherical neutral detector (SND) at the 
VEPP-2M collider in Novosibirsk. Later on, an error in software was found
and the data corrected \cite{snd2006}. We use the data on pion form factor 
from their Table I. The experimental value of the charged pion mean square
radius \cite{amendolia} is added to the fitted data set. The comparison
of the GS and our parametrization (both include also the contribution
from the $\phi$ meson) is reported on in
Table~\ref{tab_snd}. We also reproduce there the results of the updated
\cite{snd2006} fit by the SND collaboration. The formulas used in their
fitting are described in \cite{snd2005} together with three choices of the
$\rho^\prime$ and $\rho^{\prime\prime}$ parameters. Unfortunately, in the
updated paper the provided information is scarce. It is not said which
choice was used and what was the $\chi^2$ of the fit. Nevertheless, there
is a good compatibility in both $m_\rho$ and $\Gamma_\rho$ between the GS 
parametrization and that used by the SND collaboration. 

The GS and our parametrization provide equally good fits,
but the mass and width resulting from the latter are lower than and
incompatible with those from the former.

\begin{table}
\caption{\label{tab_snd}Mass and width of $\rho(770)$ from fitting the SND data.}
\begin{ruledtabular}
\begin{tabular}{lccc}
                        & GS        & ours & SND fit \cite{snd2006}\\
 \hline
 $\chi^2/NDF$        & 39.94/38    & 39.80/38        &     not given     \\
 $m_\rho$~(MeV/$c^2$)& $774.60\pm0.52$ & $764.23\pm0.43$ & $774.6\pm0.4\pm0.5$ \\
 $\Gamma_\rho$~(MeV) & $147.3\pm1.0$ & $140.29\pm0.85$ & $146.1\pm0.8\pm1.5$ \\
\end{tabular}
\end{ruledtabular}
\end{table}

\begin{figure}
\setlength \epsfxsize{8.6cm}
\epsffile{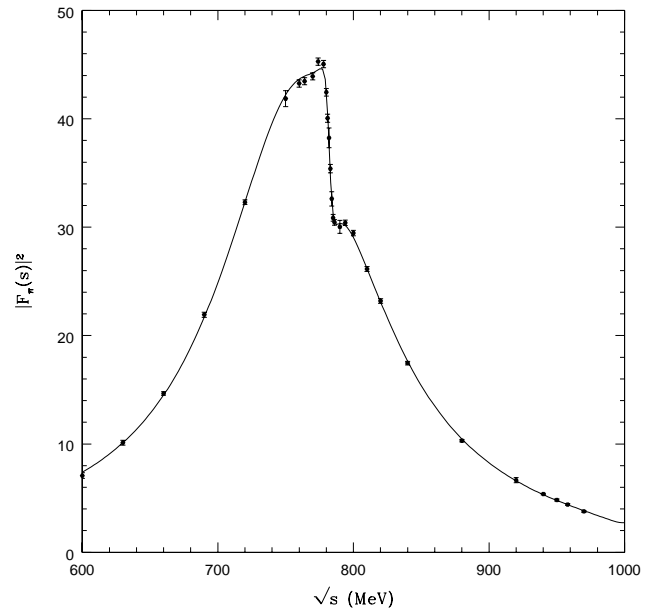}
\caption{\label{fig:snd}Comparison of our parametrization of the pion
form factor with the SND data \cite{snd2006}}
\end{figure}

\subsection{\label{sec:kloe}Using the KLOE experiment data}
The data on the cross section $\sigma(\eepp\gamma)$ at the collision
energy $W=$1.02~GeV measured with the KLOE detector at the electron-positron 
collider DA$\Phi$NE in Frascati were used in Ref.~\cite{kloe} to extract 
the square of the charged pion form factor $|F_\pi(s)|^2$ by the radiative 
return method \cite{radret}. Here, $s<W^2$ denotes the square of the
invariant mass of the $\pi^+\pi^-$ system. No attempt was made by the authors 
of \cite{kloe} to fit their form factor in order to get the $\rho$ meson
parameters. We have performed the fit using both GS and our parametrizations.
The results are summarized in Table~\ref{tab:kloe}. The $\chi^2$ of both fits
is poor. Nevertheless, the $\chi^2$ of our parametrization is about half 
that of the GS parametrization. An interesting
observation is that not only our parametrization, but also the GS one gives
a $\rho(770)$ mass substantially smaller than both Novosibirsk
experiments. As before, our parametrization yields smaller values of both 
the mass and width than GS. An attempt to improve the quality of the fits
by including the $\rho$-$\phi$ interference somewhat lowered the $\chi^2$ 
for both parametrizations, but not enough to get the confidence level into 
an acceptable range.
\begin{table}
\caption{\label{tab:kloe}Mass and width of $\rho(770)$ from fitting the KLOE data.}
\begin{ruledtabular}
\begin{tabular}{lcc}
                        &  GS            & ours \\
 \hline
 $\chi^2/NDF$           & 367.8/54       & 192.6/54      \\
 $m_\rho$~(MeV/$c^2$)   & $769.22\pm0.32$  & $761.39\pm0.25$ \\
 $\Gamma_\rho$~(MeV)    & $144.71\pm0.56$  & $140.07\pm0.66$ \\
\end{tabular}
\end{ruledtabular}
\end{table}

\begin{figure}
\setlength \epsfxsize{8.6cm}
\epsffile{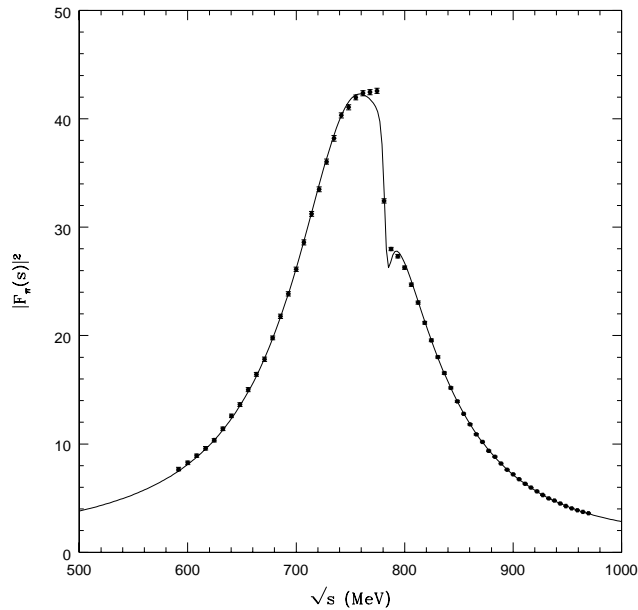}
\caption{\label{fig:kloe}Comparison of our parametrization of the pion
form factor with the KLOE data \cite{kloe}.}
\end{figure}

Figure~\ref{fig:kloe} shows that the worst discrepancy of our parametrization 
with the data is located in the $\rho$-$\omega$ region. The abrupt fall, which 
is present in Figs.~\ref{fig:cmd2} and \ref{fig:snd}, is missing. A more
detailed analysis of the latter figures showed that the steep descent
is caused by the $\rho$-$\omega$ interference term, which is very sensitive
to the exact position of the (very narrow) $\omega$ resonance. The absence 
of the interference term in Fig.~\ref{fig:kloe} suggests that the location of
the $\omega$ resonance in the KLOE data is different from the nominal one,
which we assumed when doing the fit with both GS and our parametrization.
It would mean that the energy scale in the KLOE data contains a systematic
shift, at least in the $\rho$-$\omega$ region. The KLOE
collaboration utilized the radiative return method \cite{radret}. This means
that the invariant mass of the $\pi^+\pi^-\gamma$ system (denoted as $W$
in \cite{kloe}) is fixed by the energy of the colliding $e^+$ and $e^-$. 
But the invariant mass $\sqrt{s}$ of the $\pi^+\pi^-$ subsystem has to be 
calculated from the measured momenta of the outgoing particles and is 
therefore known less precisely.

\begin{figure}
\setlength \epsfxsize{8.6cm}
\epsffile{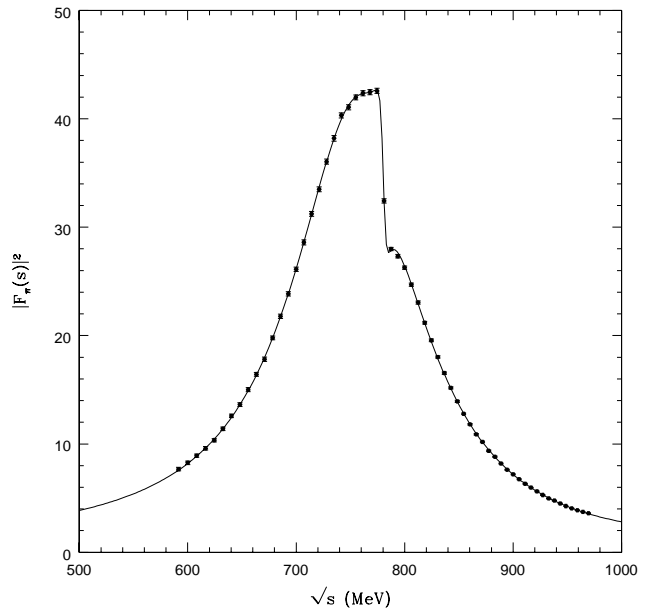}
\caption{\label{fig:kloe_omega}Comparison of our parametrization of 
the pion form factor with the KLOE data \cite{kloe} after decreasing the
$\omega(782)$ mass parameter by 1.93~MeV.}
\end{figure}

To scrutinize the possibility of a shift in experimental values of $\sqrt{s}$,
we repeated the fits allowing the $\omega$ mass also to vary. Both the
GS and our parametrization indicate that the best fit is obtained if the
$\omega$ mass is smaller than the nominal one by about 2~MeV, see
Table~\ref{tab:kloe_omega}. The drop of $\chi^2$ in both parametrizations
justifies including the $\omega$ mass among free parameters, compare
Table~\ref{tab:kloe_omega} with Table~\ref{tab:kloe}. But the quality of the
fit is very different for the two parametrizations. While the value of
$\chi^2=202.6$ for the GS parametrization is still very high (almost four 
times the number of degrees of freedom), our parametrization yields 
$\chi^2=29.6$, which means, considering $NDF=53$, the confidence level of 
0.9962. Such perfect fits are rarely encountered in the high energy physics.  
The improvement of the fit is clearly seen by naked eye if one compares 
the new parametrization (Fig.~\ref{fig:kloe_omega}) with the old one
(Fig.~\ref{fig:kloe}).

Assuming that our analysis really signifies that a systematic shift 
of $\sqrt{s}$ is present in the data, we used the difference 
$\Delta m_\omega$ between the nominal mass of the $\omega(782)$ and that
resulting from our fit to get the ``corrected" $\rho$-meson mass, see
Table~\ref{tab:kloe_omega}.
\begin{table}
\caption{\label{tab:kloe_omega}Fitting the KLOE data, the $\omega(782)$
mass allowed to vary.}
\begin{ruledtabular}
\begin{tabular}{lcc}
                      &  GS            & ours \\
 \hline
 $\chi^2/NDF$         & 202.6/53       & 29.58/53      \\
 $m_\omega$~(MeV/$c^2$)   & $780.52\pm0.33$  & $780.72\pm0.15$ \\
 $\Delta m_\omega$~(MeV/$c^2$)\footnote{Difference from $m_\omega=
(782.65\pm0.12)$~MeV \cite{pdg2006}}   & $2.13\pm0.35$  & $1.93\pm0.19$ \\
$m_\rho$~(MeV/$c^2$) & $769.45\pm0.25$  & $761.51\pm0.14$ \\
$m_\rho$~(MeV/$c^2$)\footnote{The value resulting from the fit was 
increased by $\Delta m_\omega$}   
& $771.58\pm0.60$  & $763.44\pm0.33$ \\
 $\Gamma_\rho$~(MeV)  & $146.39\pm$0.51  & $142.56\pm0.41$ \\
\end{tabular}
\end{ruledtabular}
\end{table}

Unfortunately, our interpretation of the mass shift in the
$\rho$-$\omega$ mass region as an experimental effect is
difficult to be explained in terms of a genuine miscalibration
of the KLOE drift chamber~\cite{venanzoni}.

In an effort to reveal the origin of the mysterious mass shift we 
considered the possibility that it was an artifact of a wrong 
parametrization. We combined our parametrization of the $\rho^0$
contribution \rf{piffgen} to the pion form factor, which is based on
Eq.~\rf{singledr}, with various options for other contributions, see
Sec.~\ref{sec:other}. The effect did not disappear and the value of
the $\omega$ mass shift was very stable against all changes. Moreover, 
we have found that the inclusion of the $\phi(1020)$ contribution
further improves the quality of the fit not only for our parametrization
(where the quality was already excellent), but also very significantly   
for the GS parametrization, see Tab.~\ref{tab:kloe_omega_phi}.

\begin{table}
\caption{\label{tab:kloe_omega_phi}As Tab.~\ref{tab:kloe_omega}, but
including the  $\rho$-$\phi$ interference.}
\begin{ruledtabular}
\begin{tabular}{lcc}
                      &  GS            & ours \\
 \hline
 $\chi^2/NDF$         & 58.35/51       & 25.67/51      \\
 $m_\omega$~(MeV/$c^2$)   & $780.15\pm0.17$  & $780.71\pm0.15$ \\
 $\Delta m_\omega$~(MeV/$c^2$)\footnote{Difference from $m_\omega=
(782.65\pm0.12)$~MeV \cite{pdg2006}}   & $2.50\pm0.21$  & $1.94\pm0.19$ \\
$m_\rho$~(MeV/$c^2$) & $770.29\pm0.16$  & $761.54\pm0.16$ \\
$m_\rho$~(MeV/$c^2$)\footnote{The value resulting from the fit was 
increased by $\Delta m_\omega$}   
& $772.79\pm0.35$  & $763.48\pm0.33$ \\
 $\Gamma_\rho$~(MeV)  & $151.53\pm$0.55  & $142.78\pm0.46$ \\
\end{tabular}
\end{ruledtabular}
\end{table}

\section{\label{sec:summary}Summary and Comments}
The main results of this work are summarized in Tables \ref{conflevel}, 
\ref{rhomass}, and \ref{rhowidth}. 

Table \ref{conflevel} shows how well are the particular data sets (CMD-2, SND,
KLOE) fitted by two parametrizations (GS and ours) we
considered. As a measure of the goodness-of-fit we used the confidence level
calculated from the values of $\chi^2$ and the numbers of degrees of freedom
shown in Tables \ref{tab_cmd2}--\ref{tab:kloe_omega_phi} using the CERNLIB
routine \textsc{PROB}. 
\begin{table}
\caption{\label{conflevel}Confidence level of the GS and our fits to 
various data sets}
\begin{ruledtabular}
\begin{tabular}{lcc}
   & GS & ours\\
 \hline
 CMD-2    & 0.5555  & 0.6943\\
 SND      & 0.3840  & 0.3899\\
 KLOE     & $<\!10^{-30}$  & $2\!\dek{-17}$\\
 KLOE\footnote{With the $\omega$ mass shifted} & $2\!\dek{-19}$ & 0.9962\\
 KLOE\footnote{With the $\omega$ mass shifted and $\phi$ contribution
 included} & 0.2234 & 0.9988 \\
\end{tabular}
\end{ruledtabular}
\end{table}
The agreement of both the GS and our parametrizations with both the CMD-2 and 
SND data is on the level that is considered satisfactory in the high energy
physics (based on $\chi^2/NDF\approx1$). Our parametrization gives a
slightly better confidence level than the GS one, especially for the
CMD-2 data. 

The disagreement of both parametrizations with the KLOE data is
disastrous. But the situation radically improves for our parametrization
if we allow the mass of the $\omega(782)$ to vary. Adding this free
parameter reduces the original $\chi^2=192.6$ to $\chi^2=29.6$, what 
from the statistics point of view fully justifies this step.  The confidence
level of the hypothesis that our parametrization describes correctly the
KLOE data has rocketed to 99.62\%. This is highly non-trivial because the
statistical errors of the KLOE data are very small, much smaller than those
of both sets of data from Novosibirsk. The price that has to be paid for
this excellent result is lowering the $\omega(782)$ mass by 
$\Delta m_\omega=(1.93\pm0.19)$~MeV. The physics behind this in unclear.
We originally suspected that the measured values of the mass of the 
two-pion system might contain a systematic shift, at least in the
$\rho$-$\omega$ interference region. But the correspondence with the members 
of the KLOE collaboration excluded this possibility. 

Also in the case of the GS parametrization $\chi^2$ decreased from 367.8
to 202.6 when the $\omega$ mass was considered an additional free parameter.
The required mass shift is even higher than in the case of our 
parametrization, namely $(2.13\pm0.35)$~MeV.  
Adding the $\phi$-meson contribution improves the quality of the fit 
while changing the $\omega$-mass shift negligibly, compare 
Tabs.~\ref{tab:kloe_omega} and \ref{tab:kloe_omega_phi}.

On the other hand, the
addition of the $\omega$ mass among free parameters is not required
by the CMD-2 and SND data. The drop of $\chi^2$ is less than one 
both for the GS parametrization and ours. Why the KLOE data behaves
in a different way is unclear. The issue obviously requires 
explanation. We do not have any at the moment.

Table \ref{rhomass} summarizes the results on the $\rho(770)$ mass we
have got by fitting various data
sets using the GS and our parametrization. Its salient feature is 
that for all data sets we considered (CMD-2, SND, KLOE, KLOE with the
shifted $\omega$ mass) our parametrization gives a smaller $\rho(770)$
mass than the GS one. The differences are large, they range from 7.8~MeV
for the KLOE data to 10.3~MeV for the CMD-2 data. At the same time the errors 
of the masses themselves are very small. It looks like the two columns 
concerned different quantities. And this is probably the core of the problem.
The $\rho$-meson mass is defined in both parametrizations as a parameter
in the corresponding formula for the $\rho^0$ propagator. As the formulas 
are different, so must be the definitions and the ``experimental'' values
obtained by fitting those two parametrizations. 
      
\begin{table}
\caption{\label{rhomass}The $\rho$ meson mass obtained from various data
sets using the GS and our parametrizations}
\begin{ruledtabular}
\begin{tabular}{lcc}
          & GS & ours\\
 \hline
 CMD-2    & 775.3$\pm$1.1 & 767.08$\pm$0.83\\
 SND      & 774.60$\pm$0.52 & 764.23$\pm$0.43\\
 KLOE     & 769.22$\pm$0.32 & 761.39$\pm$0.25\\
 KLOE\footnote{Accounting for the energy shift in the $\rho$-$\omega$
region} & 771.58$\pm$0.60 & 763.44$\pm$0.33\\
KLOE\footnote{Accounting for the energy shift in the $\rho$-$\omega$
region, $\phi(1020)$ included} & 772.79$\pm$0.35 & 763.48$\pm$0.33\\
\end{tabular}
\end{ruledtabular}
\end{table}

\begin{table}
\caption{\label{rhowidth}The $\rho$ meson width obtained from various data
sets using the GS and our parametrizations}
\begin{ruledtabular}
\begin{tabular}{lcc}
 $\Gamma_\rho$ & GS & ours\\
 \hline
 CMD-2    & 143.2$\pm$2.5 & 136.1$\pm$2.2\\
 SND      & 147.3$\pm$1.0 & 140.29$\pm$0.85\\
 KLOE     & 144.71$\pm$0.56 & 140.07$\pm$0.66\\
 KLOE\footnote{Fit with the energy shift in the $\rho\omega$
region} & 146.39$\pm$0.51 & 142.56$\pm$0.41\\
 KLOE\footnote{Fit with the energy shift in the $\rho\omega$
region, $\phi$ contribution included} & 151.53$\pm$0.55 & 142.78$\pm$0.46\\
\end{tabular}
\end{ruledtabular}
\end{table}

The cure of this problem is not easy. To get the ``universal" definition 
of the $\rho$-meson mass it is insufficient to declare some formula for 
the propagator (\eg, that based on the GS parametrization) as the canonical 
one. If that preferred formula is not correct, then the different processes
or even the same process measured in  different kinematic regions will
give different values of the $\rho$ meson mass even if the experiments are
ideal. In fact, in order to compare the theoretical predictions with the 
data on some process, the propagator has to be convoluted with other 
factors (``vertices") that are characteristic for that process and depend 
also on the virtual $\rho$-meson momentum squared. If the propagator is not 
correct, then the cross section or decay width formulas of various processes 
are incorrect, but each in a different way. We illustrate this point on 
a toy example in the Appendix.

It is generally believed that the only meaningful definition of the mass 
and width of a resonance, which does not have asymptotic states,
is that in terms of the position of the pole on
the second Riemann sheet of the propagator
\cite{poledefinition}. 
The pole of the propagator induces a pole in the scattering amplitude 
(at least in perturbative field theory). The same propagator enters the 
amplitudes 
of various processes, so the pole definition should be process independent.
The pole position does not change if an analytic function without pole
is added to the amplitude to describe the background present in experiments.
So the pole definition should be also background independent. In the case
of resonances for which a reliable theoretical framework exists ($W^\pm$ and
$Z^0$ bosons) is it possible to relate the mass and width defined in terms
of the pole position to the ``mass'' and ``width'' that enter as
parameters of various fitting formulas
\cite{willenbrock91,stuart91,sirlin91}. The pole-defined mass and width
are genuine physical quantities (gauge invariant and renormalization 
scheme independent in the $Z^0$ boson case). 

There have been several attempts to find the position of the $\rho^0$-meson
pole on the second sheet of the complex $s$-plane. 

Lang and Mas-Parareda
\cite{lang79} first mapped the cut $s$-plane into a unit disc with data
about the $P$-wave partial amplitude of the $\pi\pi$ scattering falling on 
the unit circle. Then they fitted the data to the product of a pole outside
the unit circle (corresponding to the second-sheet pole in the $s$ variable)
and an analytic function represented by its Taylor series. The pole
position and the coefficients in the Taylor series were parameters of the
fit. There are different ways to suppress the importance of the
higher Taylor series coefficients and make the series finite. Unfortunately,
this brings some model dependence. In \cite{lang79} the Pietarinen 
\cite{pietarinen} smoothness condition was utilized. Lang and
Mas-Parareda used five different data sets and got the mutually compatible
values of the $\rho$ mass and width, whereas the values of those quantities
determined in experimental papers were incompatible. The pole-defined
masses were always lower than the masses from the parametrizations
used be experimentalists. 
  
Bohacik and K\"{u}hnelt \cite{bohacik} used the pole-searching method
developed in \cite{nogova} on the basis of the statistical method for 
the testing of analyticity \cite{pazman}. The central objects of the
latter are the moments $Q_n$ ($n$ natural) calculated from the
data and a smooth function depending on the data errors and density.
The moments should obey the $N(0,1)$ distribution if the investigated 
data describe an analytic function. If a pole is present, the 
mean values of the moments are expressed in terms of the pole
position and residuum. The authors of \cite{bohacik} mapped the second 
sheet on the unit disc, so the assumed pole lay inside the unit circle.
They either determined the pole parameters from the moments calculated
directly from the data or multiplied the data first by
the Blaschke pole-killing factor and determined its parameters by
requiring the calculated $Q_n$ momenta be $N(0,1)$ distributed.
The resulting mass and width of the $\rho$ meson were compatible with
those of Ref.~\cite{lang79}. This approach suffers from model
dependence, which crawls in through the construction of the error
function and the choice how many and which moments are taken into
consideration \cite{pazman}. 

Dubnicka and collaborators \cite{dubnicka89,dubnicka99} constructed
two analytic models incorporating all pion form factor properties (including
pairs of complex conjugate poles on unphysical sheets, corresponding to
the $\rho$, $\rho^\prime$, and $\rho^{\prime\prime}$ resonances) and
fitted them to all accessible data. In both models, the (pole) mass of
the $\rho$ meson is much smaller than the values given by PDG
\cite{pdg2006}. The authors tried to keep the number of free parameters
as low as possible. As a consequence, the quality of their fits is
not very good, $\chi^2/NDF$ = 1.81 and 1.58 in \cite{dubnicka89} and
\cite{dubnicka99}, respectively.

The present work has not been intended as a final word in the field of the
pion form factor parametrization and extracting the resonance parameters.
We are even unable to offer unambiguous values of $m_\rho$ and
$\Gamma_\rho$. Natural candidates would be those coming from the 
best fit, which is our fit to the KLOE data. But those are hampered by
the presence of a mysterious energy shift in the $\rho$-$\omega$ mass
region. The main aim of this paper was to issue the warning that the 
question of the correct parametrizations is not simple and requires 
much attention. The resonance
parameters obtained by fitting the data are subject of large systematic
errors of the parametrizations and have often been underestimated so far.

\begin{acknowledgments}%
We are indebted to Achim Denig and Graziano Venanzoni from the
KLOE collaboration for the enlightening correspondence.
One of us (P.~L.) acknowledges useful discussions with David Kraus.
This work was supported by the Czech Ministry of Education, Youth and
Sports under contracts Nos. MSM6840770029 and MSM4781305903 and by the 
Bergen Computational Physics Laboratory in the framework of the European 
Community - Access to 
Research Infrastructure action of the Improving Human Potential Programme.
The paper was finished during the visit of one of us (P.~L.) to the
McGill University. The support from its Physics Department is gratefully
acknowledged.
\end{acknowledgments}

\appendix*
\section{A toy example}

Let us assume that three experimental groups A, B, and C measured 
the dependence of the voltage on the electric current of the same wire,
but in different intervals. Each interval contained twelve data points.
We simulated their measurements assuming the validity of the Ohm 
law with $R=1\Omega$ and the Gaussian distribution of experimental
values around the Ohm-law values with the relative error of 5\%.
The three generated data sets can be accessed at \cite{ohm_data}.
Each group fitted its data by its own formula. The formulas they chose 
can be written in a unified way as
\be
\
U=RI^\alpha,
\ee
with $\alpha=$~0.95, 1.05, and 0.90 for the group A, B, and C, respectively.
The results of the fits are shown in Table~\ref{fits1}. The quality of all
fits is acceptable according to the high-energy physics standards and the
resulting parameters $R$ have rather small errors. But the $R$s of the
different groups are incompatible.
\begin{table}
\caption{\label{fits1}Each group fits its data with its own formula}
\begin{ruledtabular}
\begin{tabular}{cccr}
Group& $\alpha$ & $R$ & $\chi^2$ \\
 \hline
 A  & 0.95 & $1.096\pm 0.016$ & 11.86 \\
 B  & 1.05 & $0.866\pm 0.013$ & 10.26 \\
 C  & 0.90 & $1.369\pm 0.020$ &  4.88 \\
\end{tabular}
\end{ruledtabular}
\end{table}

Then the groups realized that problem lies in different formulas they
used and agreed to adopt the formula of the B group ($\alpha=1.05$)
as the common one. They again got the good fits and better agreement
among their $R$ parameters, but some inconsistency, especially between
groups A and C, remained, see Table~\ref{fits2}.
\begin{table}
\caption{\label{fits2}Each group fits its data with the common formula
($\alpha=1.05$)}
\begin{ruledtabular}
\begin{tabular}{ccr}
Group & $R$ & $\chi^2$ \\
 \hline
 A  &  $0.928\pm 0.013$ & 10.47 \\
 B  &  $0.866\pm 0.013$ & 10.26 \\
 C  &  $0.854\pm 0.012$ &  6.10 \\
\end{tabular}
\end{ruledtabular}
\end{table}

Only after the groups learnt about the correct formula and decided to use 
common $\alpha=1$, they have obtained both good fits and compatible results, 
as it is witnessed in Table~\ref{fits3}.
\begin{table}
\caption{\label{fits3}Each group fits its data with the correct formula
($\alpha=1$)}
\begin{ruledtabular}
\begin{tabular}{ccr}
Group & $R$ & $\chi^2$ \\
 \hline
 A  &  $1.011\pm 0.015$ & 4.64 \\
 B  &  $0.987\pm 0.014$ & 8.12 \\
 C  &  $0.999\pm 0.014$ & 5.18 \\
\end{tabular}
\end{ruledtabular}
\end{table}

\end{document}